\documentclass[doublecol]{epl2} 
\usepackage[utf8]{inputenc}
\usepackage[english]{babel}
\usepackage{amsmath}
\usepackage{amssymb}
\usepackage{subfigure}
\usepackage{amsfonts}
\usepackage{amssymb}
\usepackage{graphicx}
\newcommand{\be}{\begin{equation}}
\newcommand{\ee}{\end{equation}}
\newcommand{\bea}{\begin{eqnarray}}
\newcommand{\eea}{\end{eqnarray}}

\newcommand{\1}{\chi_{{}_{{}_1}}}
\newcommand{\2}{\chi_{{}_{{}_2}}}
\newcommand{\3}{\chi_{{}_{{}_3}}}
\newcommand{\jj}{\chi_{{}_{{}_j}}}
\newcommand{\4}{\chi_{{}_{{}_{2,\,3}}}}
\newcommand{\6}{\chi_{{}_{{}_{1,\,2}}}}
\newcommand{\5}{\chi_{{}_{{}_{1,\,2,\,3}}}}

\catcode`@=12
\def\la{\mathrel{\mathchoice {\vcenter{\offinterlineskip\halign{\hfil
$\displaystyle##$\hfil\cr<\cr\sim\cr}}}
{\vcenter{\offinterlineskip\halign{\hfil$\textstyle##$\hfil\cr<\cr\sim\cr}}}
{\vcenter{\offinterlineskip\halign{\hfil$\scriptstyle##$\hfil\cr<\cr\sim\cr}}}
{\vcenter{\offinterlineskip\halign{\hfil$\scriptscriptstyle##$\hfil\cr<\cr\sim
\cr}}}}}

\title{Dwarf Galaxy $\gamma$-excess and 3.55 keV X-ray Line In A
Nonthermal Dark Matter Model}
\author{Anirban Biswas \inst{1}\footnote{Present address:
Harish-Chandra Research Institute, Chhatnag Road, Jhusi, Allahabad, 211019, India},
Debasish Majumdar \inst{1}, 
Probir Roy \inst{2}}
\institute{\inst{1} Astroparticle Physics and Cosmology Division,
Saha Institute of Nuclear Physics, Kolkata 700064, India\\
\inst{2} Centre for Astroparticle Physics and Space science,
Bose Institute, Kolkata 700091, India}
\abstract{Recent data from Reticulum II (RetII) require
the energy range of the FermiLAT $\gamma$-excess to be $\sim$
$2-10$ GeV. We adjust our unified nonthermal Dark Matter (DM) model
to accommodate this. We have two extra scalars beyond
the Standard Model to also explain 3.55 keV X-ray line. Now the
mass of the heavier of them has to be increased to lie around
250 GeV, while that of the lighter one remains
at 7.1 keV. This requires a new seed mechanism for the $\gamma$-excess
and new Boltzmann equations for the generation of the DM relic
density. All concerned data for RetII and the X-ray line can now be
fitted well and consistency with other indirect limits attained.} 
\begin{document} \maketitle
The endeavour for the detection of Dark Matter (DM) is increasingly gaining
momentum. Gamma-ray signals from the FermiLAT experiment have attracted much
attention [\cite{Goodenough:2009gk}-\cite{Geringer-Sameth:2015lua}]. 
These cannot be explained by the known astrophysical processes. On the
other hand, their DM origin has been a topic of debate
[\cite{Geringer-Sameth:2015lua}-\cite{Banik:2015aya}].
One possibility is the decay/self-annihilation of DM particles
clustered around massive gravitating bodies, e.g. the Galactic
Centre (GC) or dwarf galaxies. Separately, an X-ray line of energy
3.55 keV has been reported [\cite{Bulbul:2014sua}, \cite{Boyarsky:2014jta}]
by the XMM Newton observatory by use of a data set obtained
from Andromeda and 73 other galaxy clusters including Perseus.
An astrophysical explanation \cite{Phillips:2015wla} of this line,
though possible, is beset \cite{Iakubovskyi:2015wma} with uncertainties in the
potassium abundance in the target. Thus a DM origin of the X-ray line remains
a viable possibility and could be from decaying [\cite{Borah:2015rla},
\cite{Arcadi:2014dca}] annihilating \cite{Baek:2014poa}
or excited \cite{Okada:2014zea} DM. It would be a worthwhile effort to construct a
unified DM model for these two phenomena.

Data from the dwarf spheroidal galaxy RetII \cite{Geringer-Sameth:2015lua}
suggest an upward shift in the earlier claimed [\cite{Vitale:2009hr}-\cite{Daylan:2014rsa}]
energy range of the FermiLAT $\gamma$-excess to $2-10$ GeV. The high
galactic latitude of RetII makes its $\gamma$-emission  relatively
free from complicated backgrounds. This higher range is what we adopt
here. That requires a modification in our 2-component nonthermal DM model
\cite{Biswas:2015sva}, proposed earlier to explain both the $\gamma$-excess
and the X-ray line. In our model the fields describing
DM have tiny couplings with Standard Model (SM) fields.
As a result, the DM particles are produced nonthermally and
they are unable to thermalise later. Two extra
electroweak (EW) singlet scalar fields $S_{{}_{2,\,3}}$
are introduced. These and the SU(2)$_{\rm L}$ doublet Higgs field
$H$ comprise the scalar sector. Inter-mixing among them leads
to three physical particles $\5$ with $M_{\1} \sim 125$ GeV, $\2$
and $\3$ (with $M_{\3} \sim 7$ keV) having tiny
mixing angles between them. The decays $\3\rightarrow \gamma \gamma$
and $\2\rightarrow {\rm b} \bar{\rm b}$ (with the ${\rm b}$'s emitting
neutral pions via hadronisation) respectively account for the X-ray line
and $\gamma$-excess. Relic DM, a mixture of $\2$ and $\3$, forms
after EW symmetry breaking through the processes $\1\rightarrow
\4 \4$, $W^+ W^-\rightarrow \4 \4$, $ZZ \rightarrow \4 \4$,
$t\bar{t}\rightarrow\4\4$, $\1\1\rightarrow\4\4$.

An important feature here is the sensitive link between
$M_{\2}$ and the energy spectrum of the $\gamma$-excess.
Indeed, we need $M_{\2}$ in the ballpark of 250 GeV
to fit the increased energy range of this excess.
As shown numerically later, too small a magnitude of $M_{\2}$,
as compared with this ballpark value, would
unacceptably shift the energy spectrum of the $\gamma$-excess
to a lower range. On the other hand, too large a mass of
$\2$ would inhibit its pair production which took
place after the EW phase transition ($T_W \sim 153$ GeV \cite{Babu:2014pxa}).
Now the decay $\1\rightarrow\2\2$ is disallowed and $\2$'s are produced in the
early Universe from the pair annihilation of SM fermions
and gauge bosons. Moreover, the decay $\2 \rightarrow W^+ W^-$
is now allowed. The strength of the $\2 W^+ W^-$ ($\2 {\rm b} \bar{\rm b}$) coupling
is proportional to $M^2_W/v$ ($g m_{\rm b}/M_W$), $g$ being the SU(2)$_{\rm L}$
gauge coupling. Consequently, the $\2\rightarrow W^+ W^-$ decay channel 
becomes the dominant contribution to our seed mechanism
for the $\gamma$-excess.

Let us recount the salient features of our model. The stability
of all scalar fields is ensured by the discrete symmetry 
$\mathbb{Z}_2 \times \mathbb{Z}^{\prime}_2$. With respect to this,
$S_{{}_2}$ and $S_{{}_3}$ have charges (-1, 1) and
(1, -1) respectively, while those of all other SM fields
are (1, 1). The scalar potential for the Higgs portal is given by
$V=V_0 + V^{\prime}$ where

\begin{eqnarray}
V_0(H, S_{{}_2}, S_{{}_3}) &=&
\kappa_{{}_1} (H^{\dagger}H-
\frac{1}{2} v^2)^2 + \frac{1}{4} \kappa_{{}_2} S_{{}_2}^4
\nonumber \\&& 
+\frac{1}{4} \kappa_{{}_3} (S_{{}_3}^2-u^2)^2 
+ \frac{1}{2} \rho_{{}_2}^2 S_{{}_2}^2
\nonumber \\&& 
+ \lambda_{12} (H^{\dagger}H) S_{{}_2}^2 
+ \lambda_{23} S_{{}_2}^2 S_{{}_3}^2 \nonumber \\
&& +~\lambda_{13}(H^{\dagger}H-\frac{1}{2} v^2)
(S_{{}_3}^2 - u^2) \,\,,
\nonumber \\
V^\prime(S_{{}_2}, S_{{}_3}) &=& \alpha\,S_{{}_2} S_{{}_3} \,\,.
\label{potential}
\end{eqnarray}
Terms such as $(H^\dagger H)S_{{}_2} S_{{}_3}$,
$S_{{}_2}^3 S_{{}_3}$, $S_{{}_2} S_{{}_3}^3$ are excluded
by the assumed symmetry. The ``small" $V^{\prime}$ term softly and
explicitly breaks the $\mathbb{Z}_2 \times \mathbb{Z}^{\prime}_2$
invariance down to that of $\mathbb{Z}^{\prime \prime}_2$ under which
$S_{{}_{2,\,3}}$ are odd and the rest are even. This $\mathbb{Z}^{\prime \prime}_2$
is spontaneously broken by the VEV $u$ ($2\,{\rm MeV}<u \leq 10\,{\rm MeV}$)\footnote{See
right panel of Fig.\ref{xray-parameter-plot}
and the related discussion.} of $S_{{}_3}$. In (\ref{potential})
$v= \langle\rm Re\,H^0 \rangle$, $H^0$ being the neutral member
of the doublet $H$, while the coupling constants
$\kappa_{1,2,3}$, $\rho_2$, $\lambda_{12}$, $\lambda_{23}$
and $\lambda_{13}$ obey certain stability conditions
detailed in Ref. \cite{Biswas:2015sva}. Domain wall formation from
the restoration of $\mathbb{Z}^{\prime\prime}_2$ at a high
temperature can also be shown to be inconsequential \cite{Babu:2014pxa}.

\begin{figure}[h!]
\centering
\includegraphics[height=8cm,width=5.5cm,angle=-90]{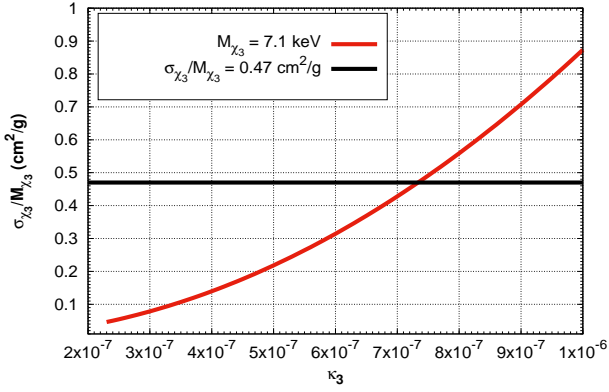}
\caption{Variation of ${\sigma_{\3}}/{M_{\3}}$ with $\kappa_3$.}
\label{figselfint}
\end{figure}

The physical scalar fields are 
$s_1 = \sqrt{2}{\rm Re}\,H^0-v$, $s_{{}_2} = S_{{}_2}$ and
$s_{{}_3} = S_{{}_3}-u$ with their squared mass matrix
\begin{eqnarray}
\mathcal{M}^2=\left( \begin{array}{ccc}
2 \kappa_{{}_1} v^2 & 0 & 2\lambda_{13}u v\\
0 & \rho_{{}_2}^2 + \lambda_{12} v^2 + 2\lambda_{23} u^2 & \alpha \\
2\lambda_{13}u v  & \alpha & 2 \kappa_{{}_3} u^2 \\
\end{array}\right) \,. 
\label{mass-matrix}
\end{eqnarray} 
The eigenvalues of (\ref{mass-matrix}) are $M^2_{\5}$ with respective
eigenstate fields $\5$. The latter are linearly related to
$s_{{}_{1,\,2,\,3}}$ via the mixing angles $\theta_{12}$,
$\theta_{23}$ and $\theta_{13}$. 
These angles are quite tiny
because two of them come from symmetry breaking also
owing to the smallness of the $\lambda$'s.
$\theta_{23}$ is a pure $\mathbb{Z}_2 \times \mathbb{Z}^{\prime}_2$
symmetry breaking parameter controlled by $\alpha$ which is
chosen to be $\sim$ ($10^{-9}-10^{-8}$) GeV$^2$ while $\theta_{12}$
is generated by an interplay of $\alpha$ and $\lambda_{13}$ which
has been taken $\sim 10^{-9}$. The last mixing angle $\theta_{13}$
arises from the spontaneous breakdown of the $\mathbb{Z}^{\prime\prime}_2$
symmetry driven by $\lambda_{13}$. From a UV perspective the smallness
of $\alpha$ and the $\lambda$'s could be due to a presumed hidden tree
level symmetry broken by radiative loop corrections.    
We further choose $M_{\1}=125.5$ GeV
$M_{\2}\sim250\,{\rm GeV}$ and $M_{\3}=7.1$ keV.
The last choice is consistent with an $\mathcal{O}$ (MeV) $u$ provided
$2\times 10^{-7} < \kappa_3 < 4\times 10^{-6}$.
However, the upper bound is further restricted to $\sim7\times 10^{-7}$
if we take the $\3$ self-interaction cross section $\sigma_{\3}$ divided by
its mass $M_{\3}$ to be less than 0.47 cm$^2/g$ from collisions between
different galaxy clusters \cite{science}, cf. Fig.\ref{figselfint}. The corresponding
quantity for $\2$ is too small to make any difference.
\begin{figure}[h!]
\includegraphics[height=1.5cm,width=2.75cm]{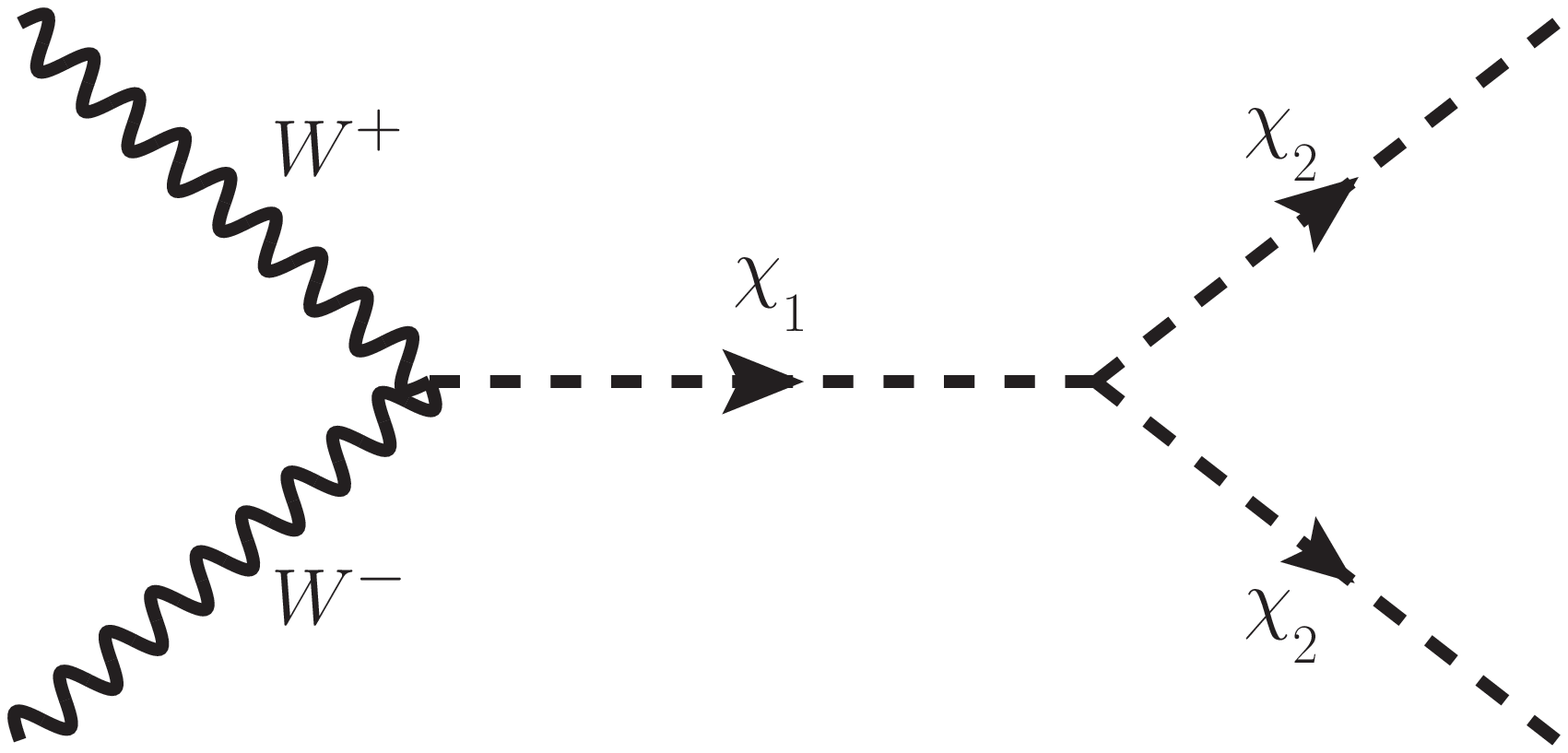}
\includegraphics[height=1.5cm,width=2.75cm]{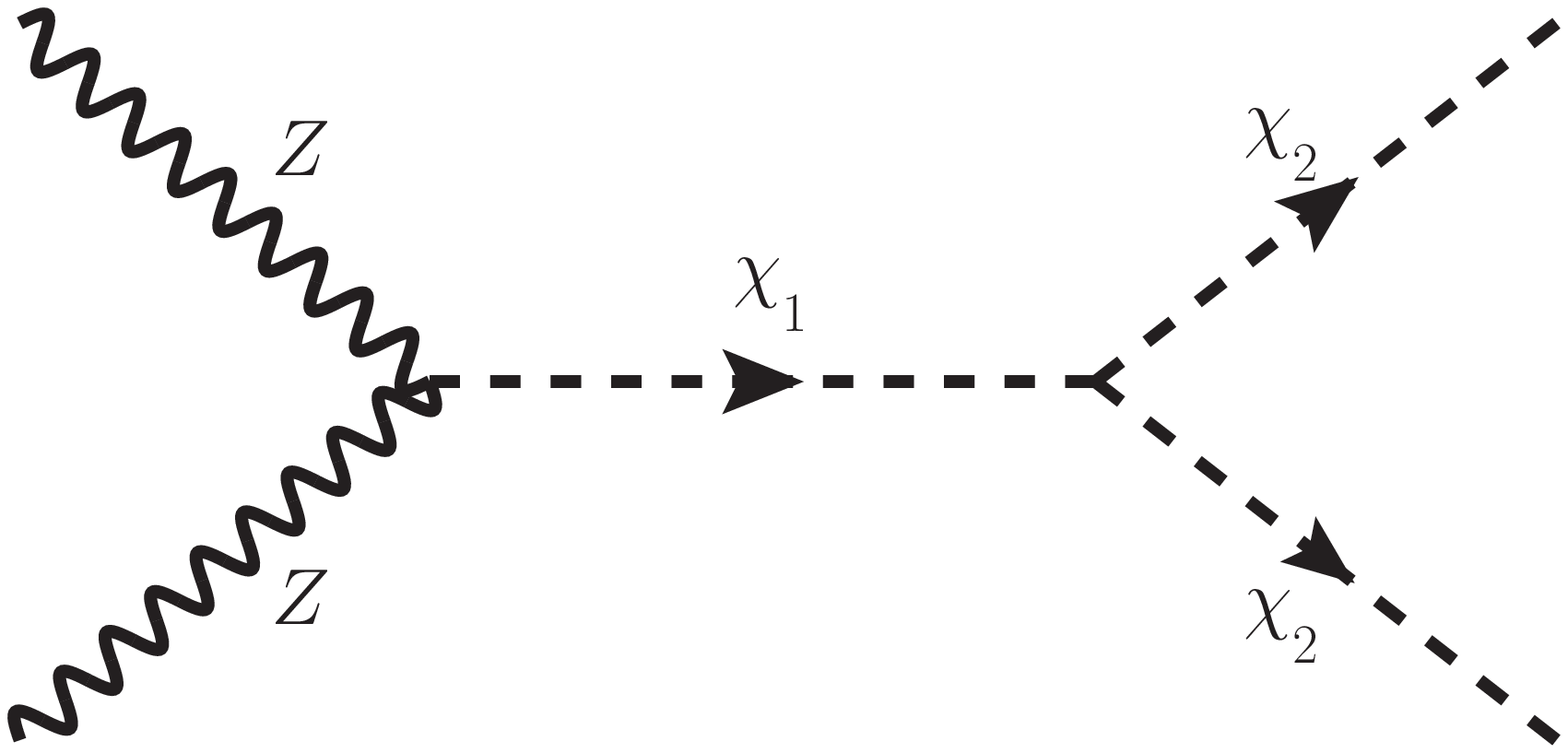}
\includegraphics[height=1.5cm,width=2.75cm]{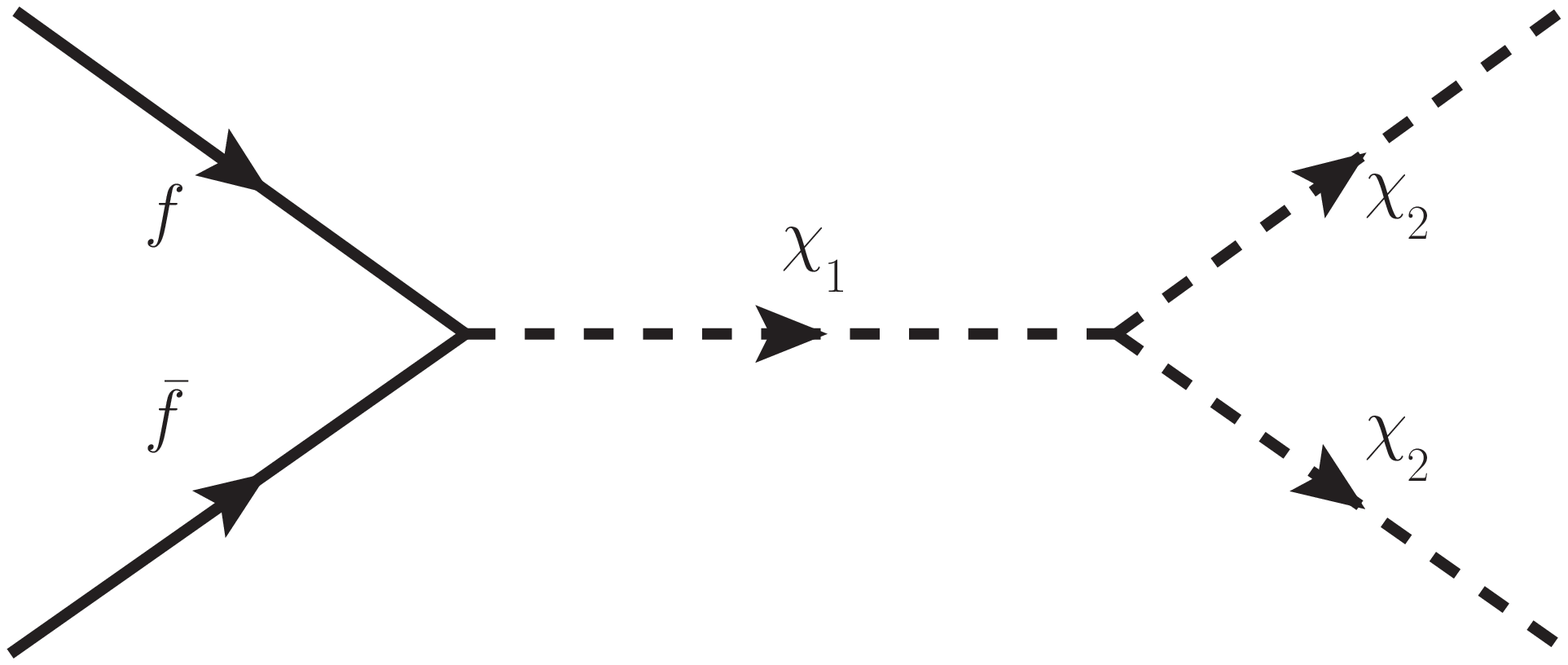}\\
\includegraphics[height=1.5cm,width=2.75cm]{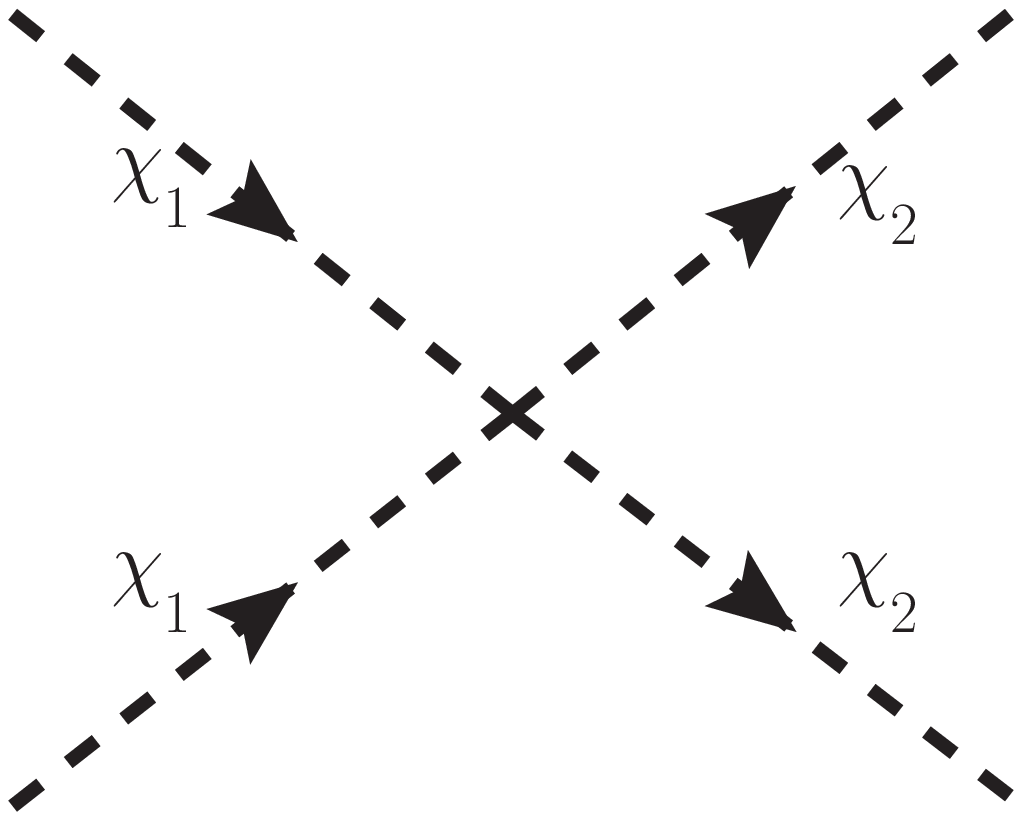}
\includegraphics[height=1.5cm,width=2.75cm]{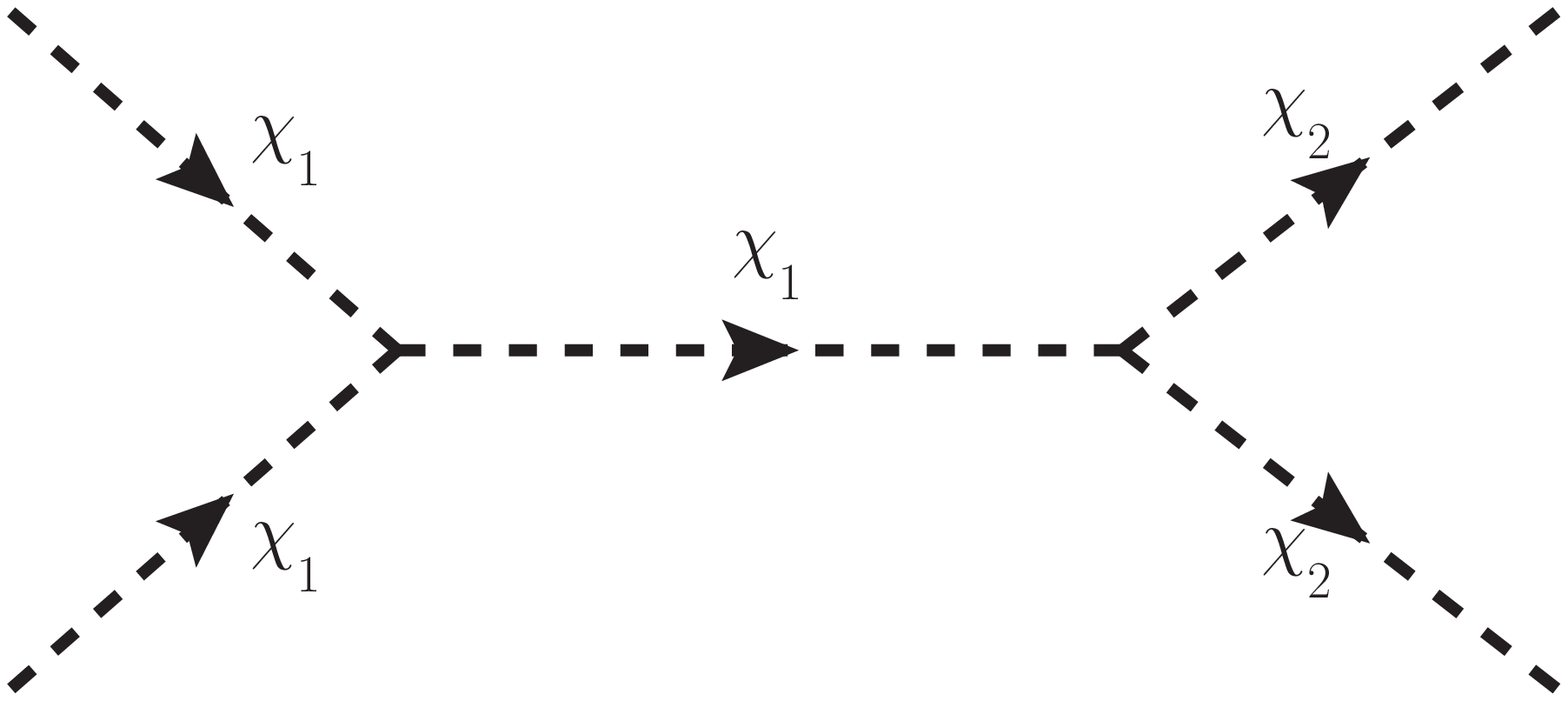}
\includegraphics[height=1.5cm,width=2.75cm]{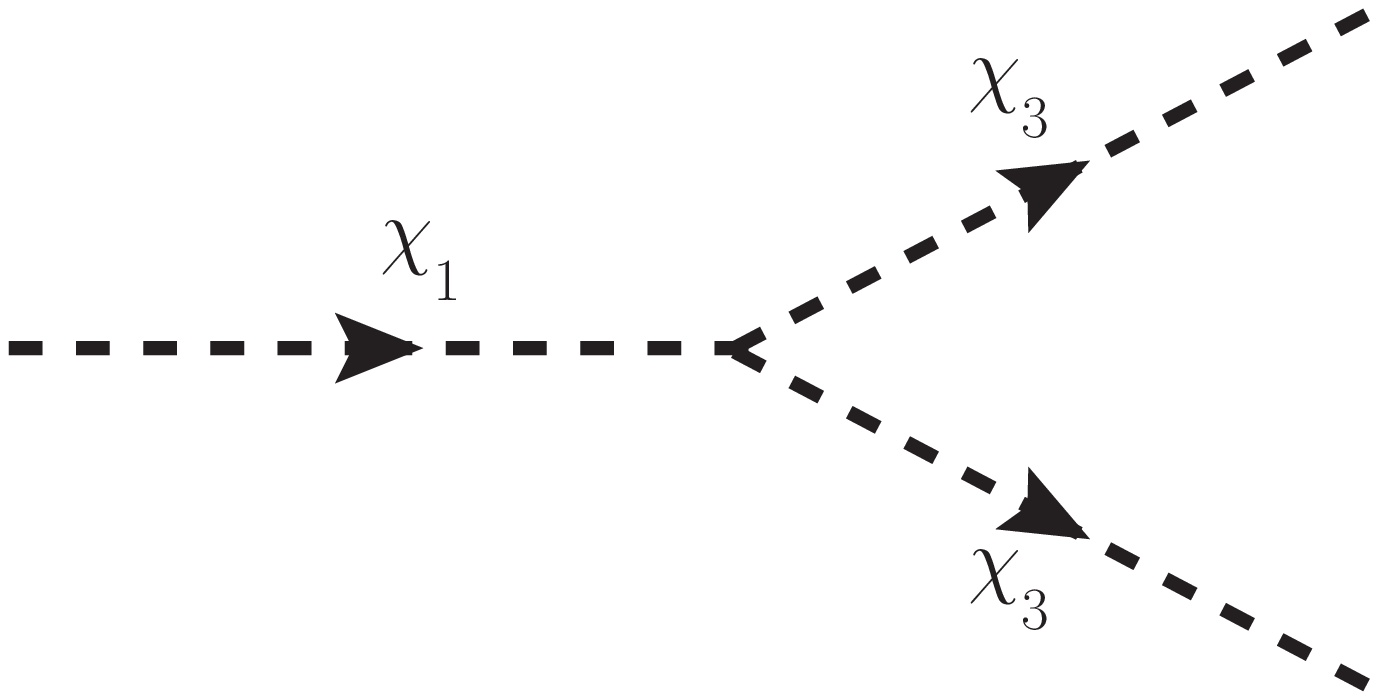}\\
\includegraphics[height=1.5cm,width=2.75cm]{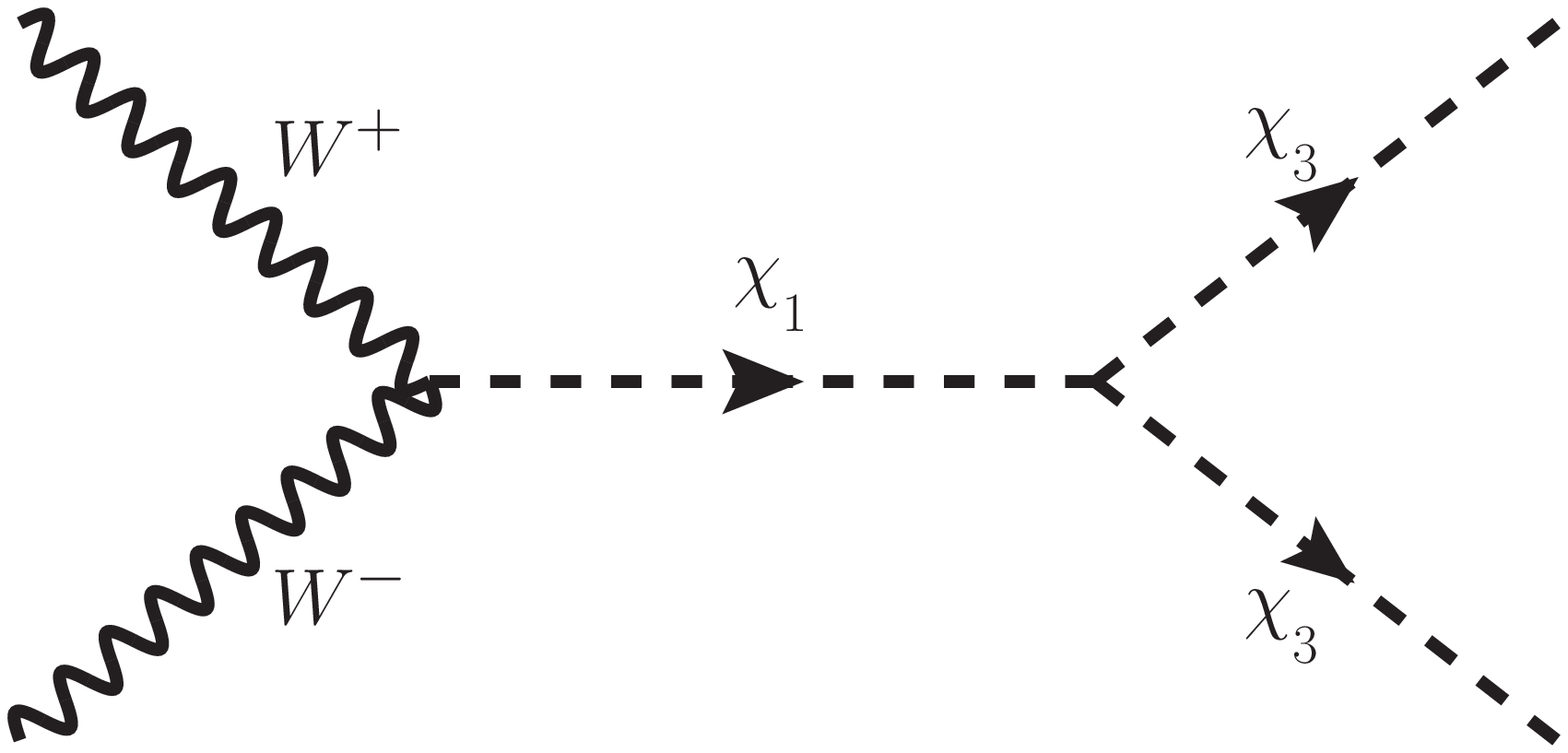}
\includegraphics[height=1.5cm,width=2.75cm]{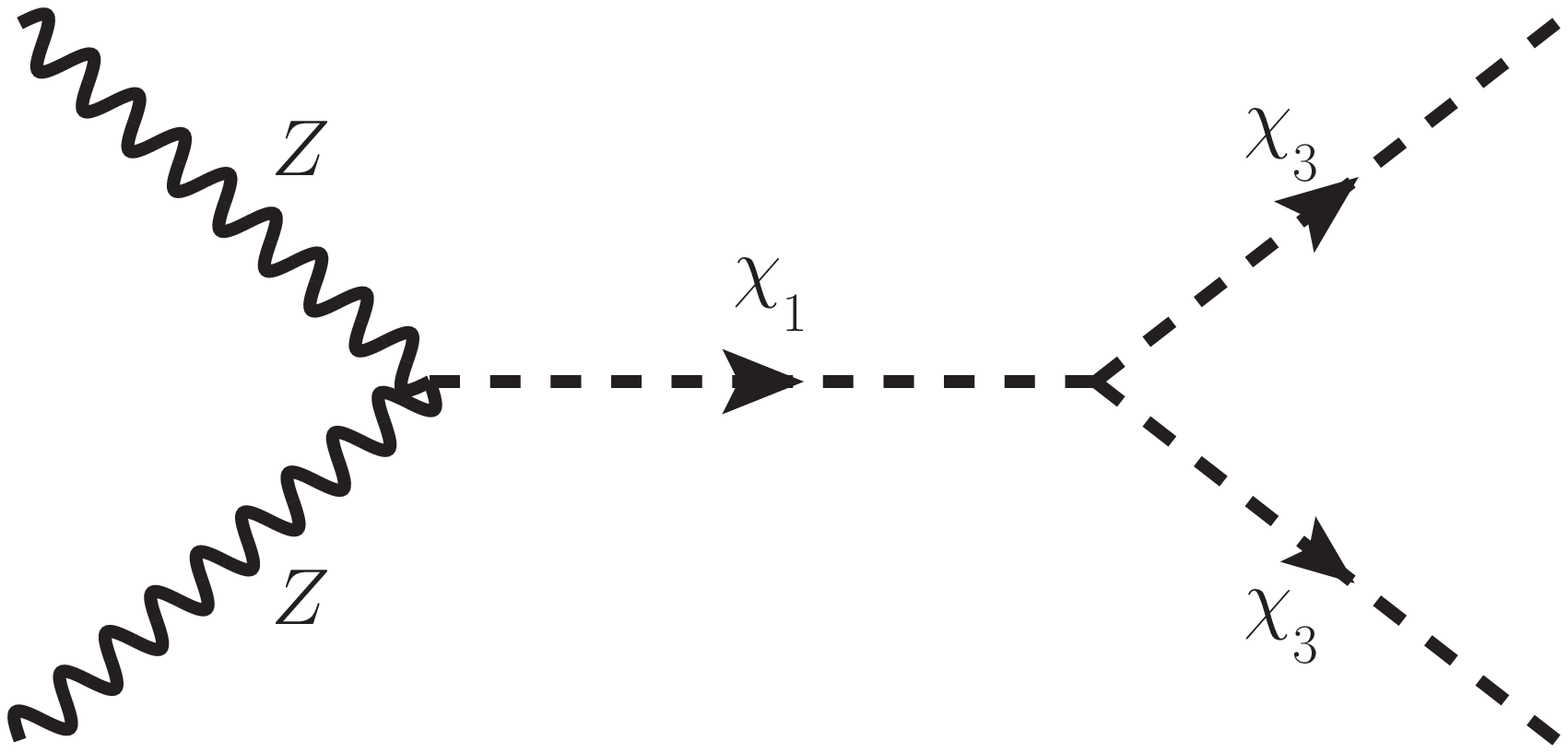}
\includegraphics[height=1.5cm,width=2.75cm]{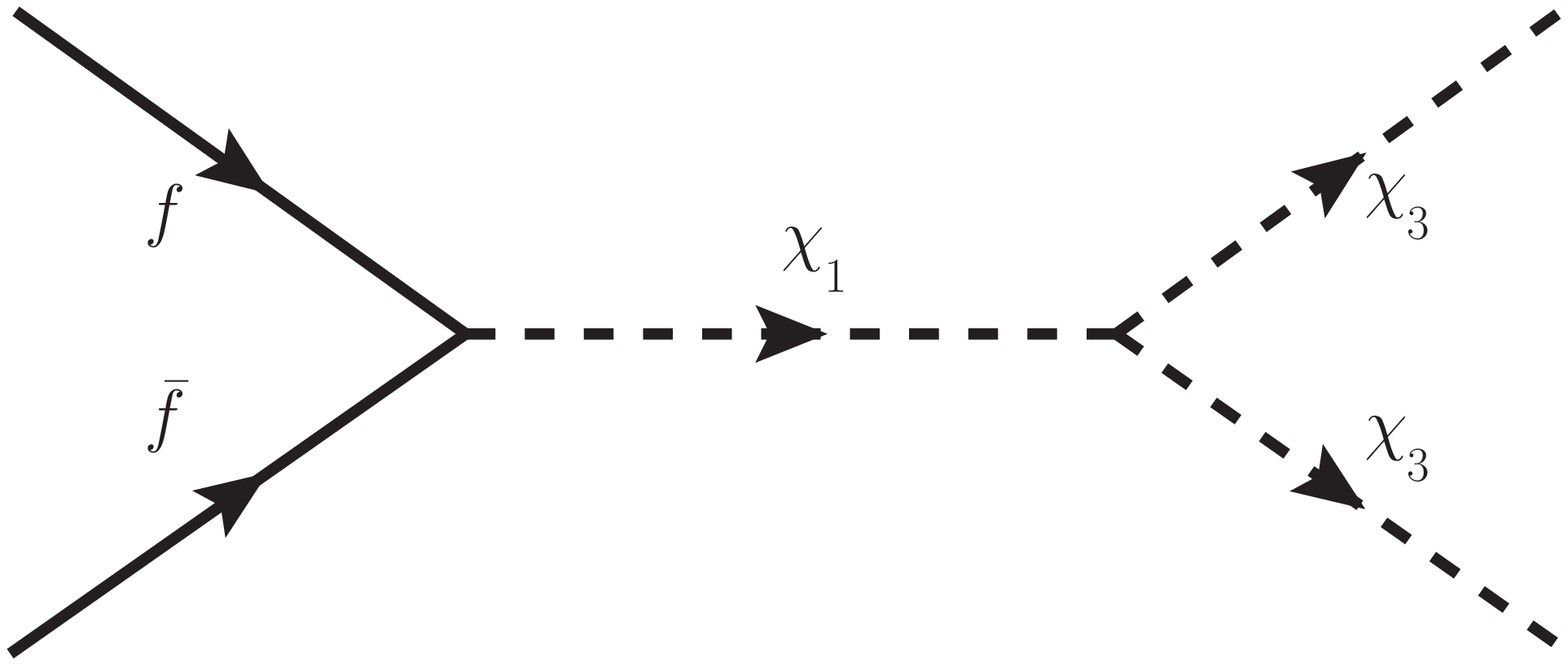}\\
\includegraphics[height=1.5cm,width=2.75cm]{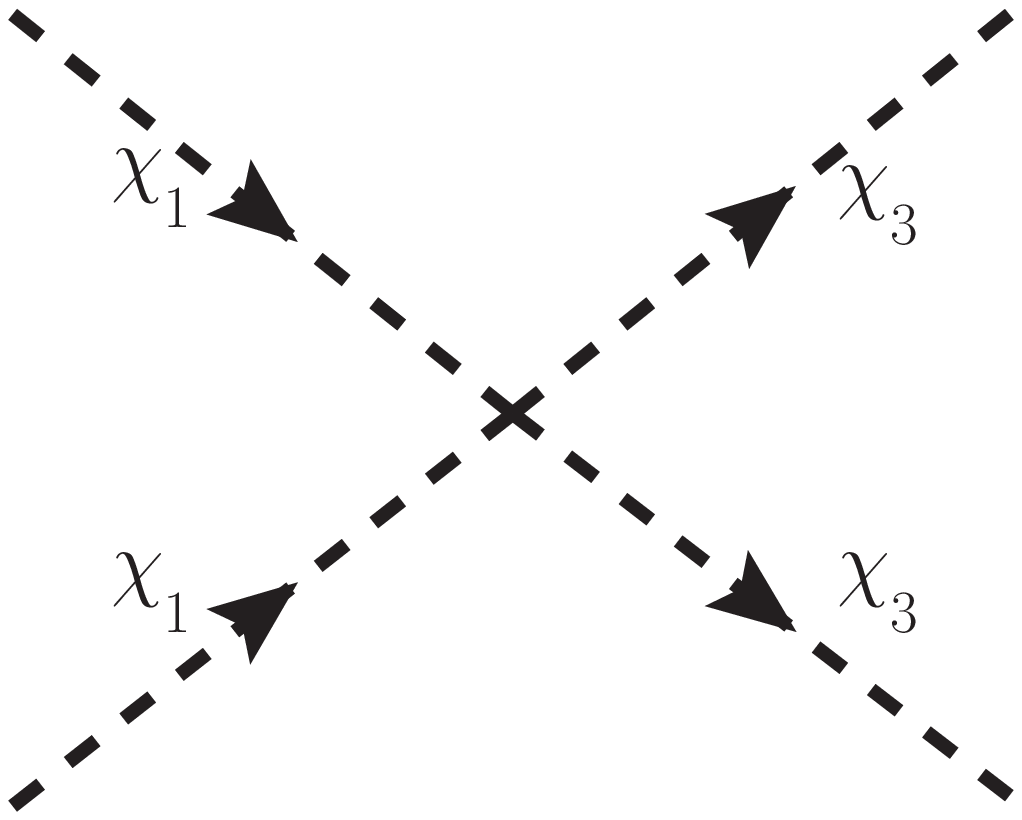}
\includegraphics[height=1.5cm,width=2.75cm]{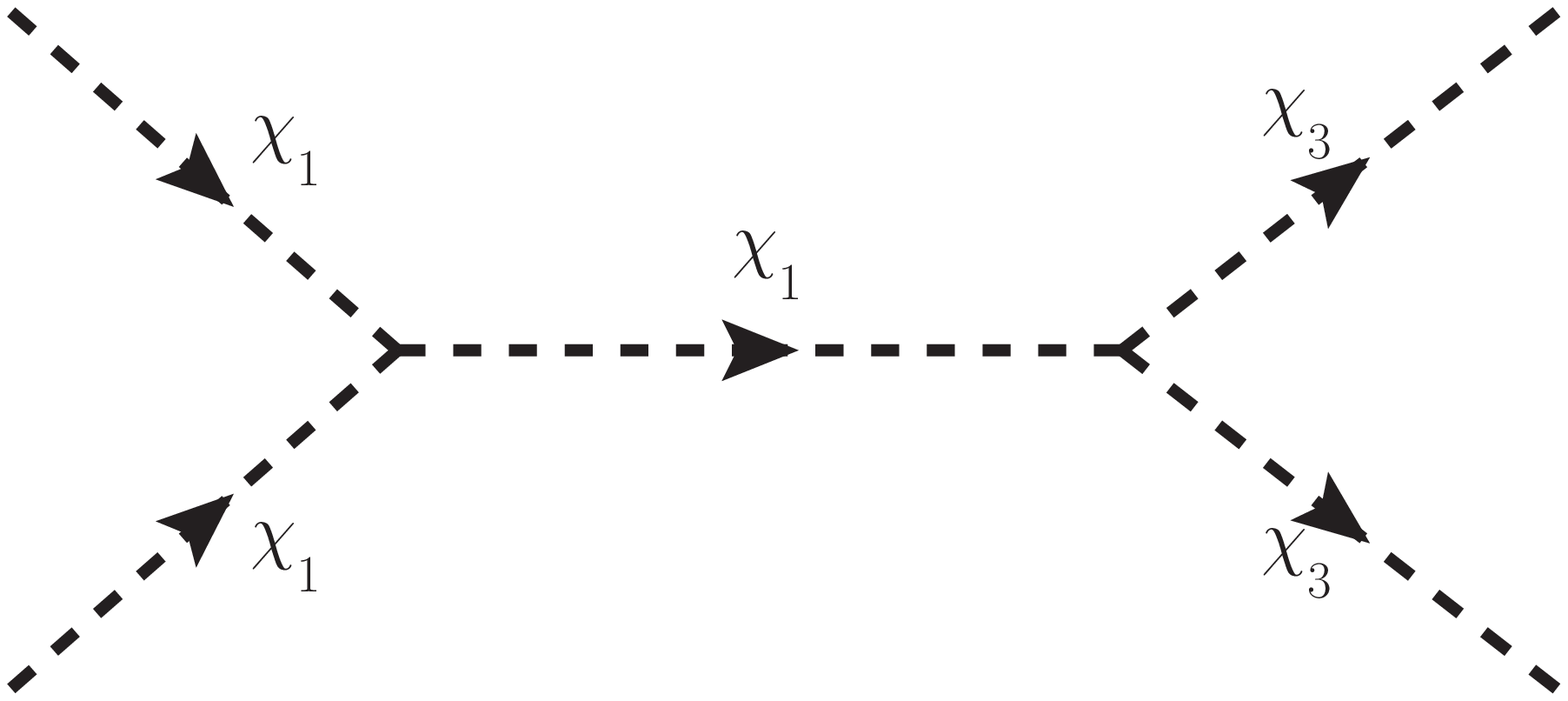}
\includegraphics[height=1.5cm,width=2.75cm]{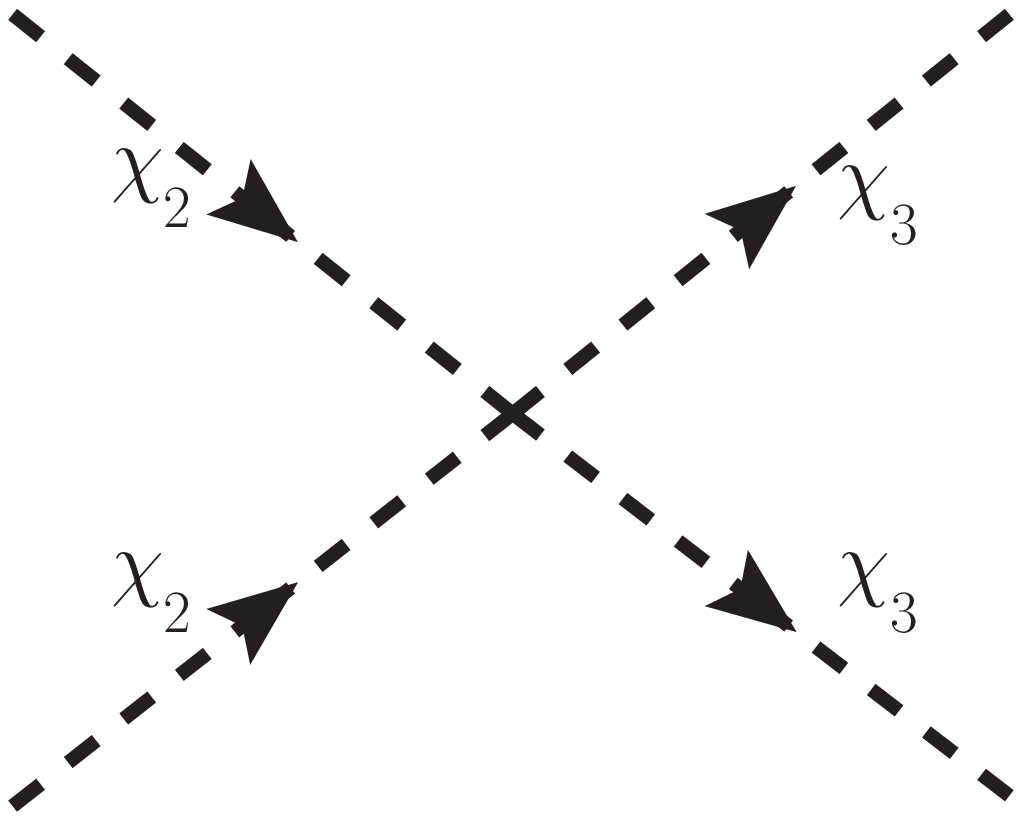}\\
\includegraphics[height=1.5cm,width=2.75cm]{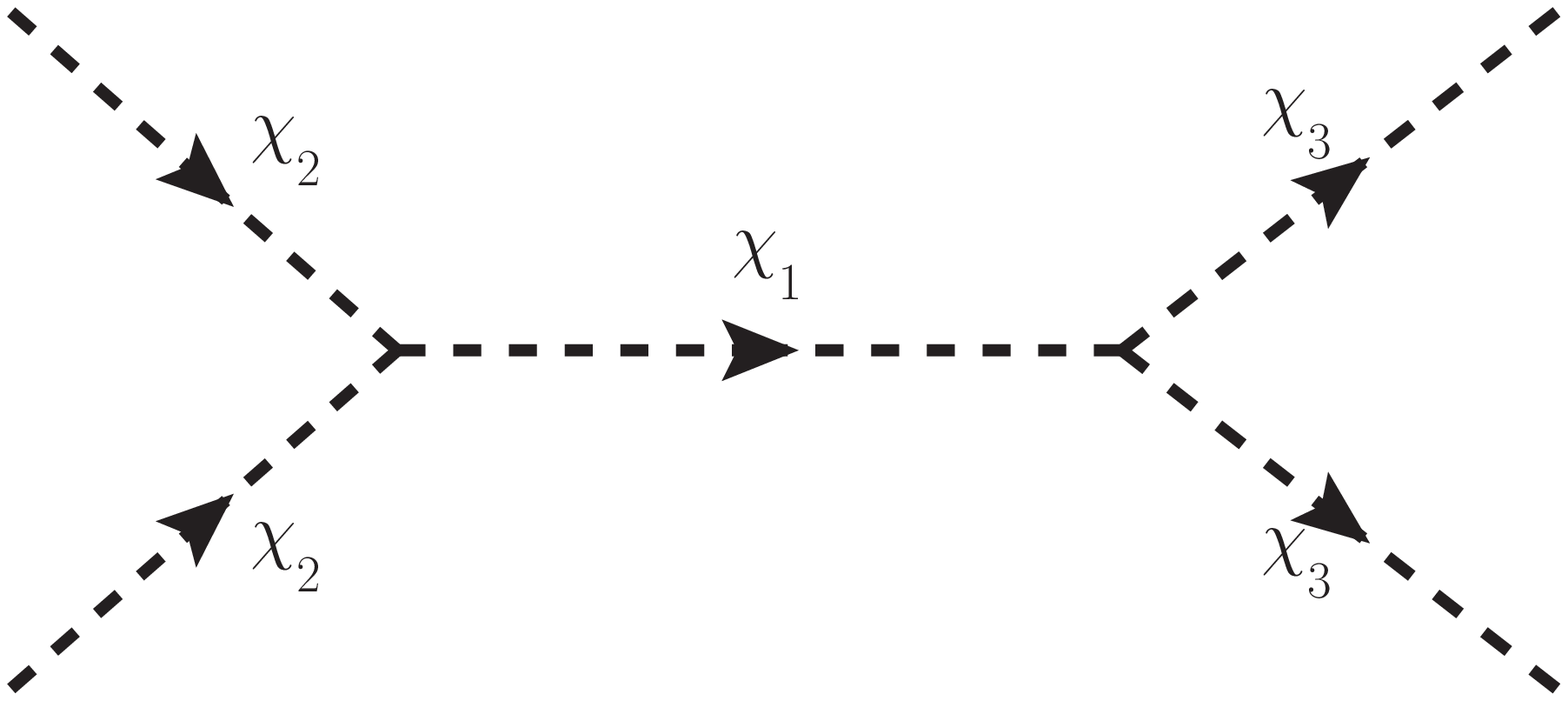}
\caption{Feynman diagrams for dominant production channels of
both the dark matter components $\chi_{{}_{{}_2}}$ and
$\chi_{{}_{{}_3}}$.}
\label{fynman-dia}
\end{figure}

Both $\2$ and $\3$ got produced nonthermally in the early Universe
but only after EW symmetry breaking. Thereafter, the self-annihilation of
$W$, $Z$, $\1$ and $t$ (see Fig.\ref{fynman-dia}) acted as
primary sources of DM particles $\4$.
The decay $\1\rightarrow \3 \3$ also contributed. Let $Y_{\jj}$ be the
comoving number density (= actual number density $\div$ entropy density of the Universe)
of $\jj$. It is given as a function of $z\equiv\frac{M_{\1}}{T}$ by a set of
two coupled Boltzmann equations. The latter involve the thermally averaged
decay width $\langle\Gamma_{\1\rightarrow\3 \3}\rangle$ as well as
the pair-production cross section times the relative velocity of collision  
${\langle \sigma {\rm v}\rangle}_{x \bar{x} \rightarrow \4\4}$ for
$x = W$, $Z$, $f$ and $\6$. Details appear in Ref. \cite{Biswas:2015sva}
and will not be repeated here. The only change is that
the decay $\1\rightarrow \2 \2$ is disallowed now. Thus, while
the equation for $\frac{d Y_{\3}}{d z}$ is unchanged, that for
$\frac{d Y_{\2}}{d z}$ is changed to
\begin{eqnarray}
&& \!\!\!\!\!\!\!\!\!\!\!\!\frac{dY_{\2}}{dz} = 
-~\frac{4\pi^2}{45\times1.66} 
M_{pl}\,M_{\chi_{{}_{{}_1}}}\sqrt{g_{\star}(T)}\,\,{z^{-2}}\,\,\times
\nonumber \\
&& \!\!\!\!\!\!\!\!\!\!\!\! \left(\sum_{a}\,({{Y}^2_{\2}}-{Y^{eq}_{a}}\,^2)
\,\langle {\sigma {\rm v}} \rangle_{a\bar{a}\rightarrow \2 \2} \,
+~ Y^2_{\2}\,\,\langle {\sigma {\rm v}} \rangle_{\2\2\rightarrow
\3 \3}\right) \nonumber \\
\label{boltz-eq2} 
\end{eqnarray} 
with $a = W$, $Z$, $f$ and $\1$. Further, the DM relic
density is given (for $j=2$, $3$) by
\begin{eqnarray}
\Omega_{\jj} h^2 = 2.755\times 10^8
\,(M_{\jj}/{\rm GeV})\,
Y_{\jj}(z_0) \,,
\end{eqnarray}
where $z_0\equiv M_{\1}/{T_0}$, $T_0$ being the present temperature
of the Universe. 
\begin{figure}[h!]
\centering
\subfigure[$\Omega_{\chi_{{}_{{}_2}}} > \Omega_{\chi_{{}_{{}_3}}}$]
{\includegraphics[height=4.15cm,width=4.5cm,angle=-90]{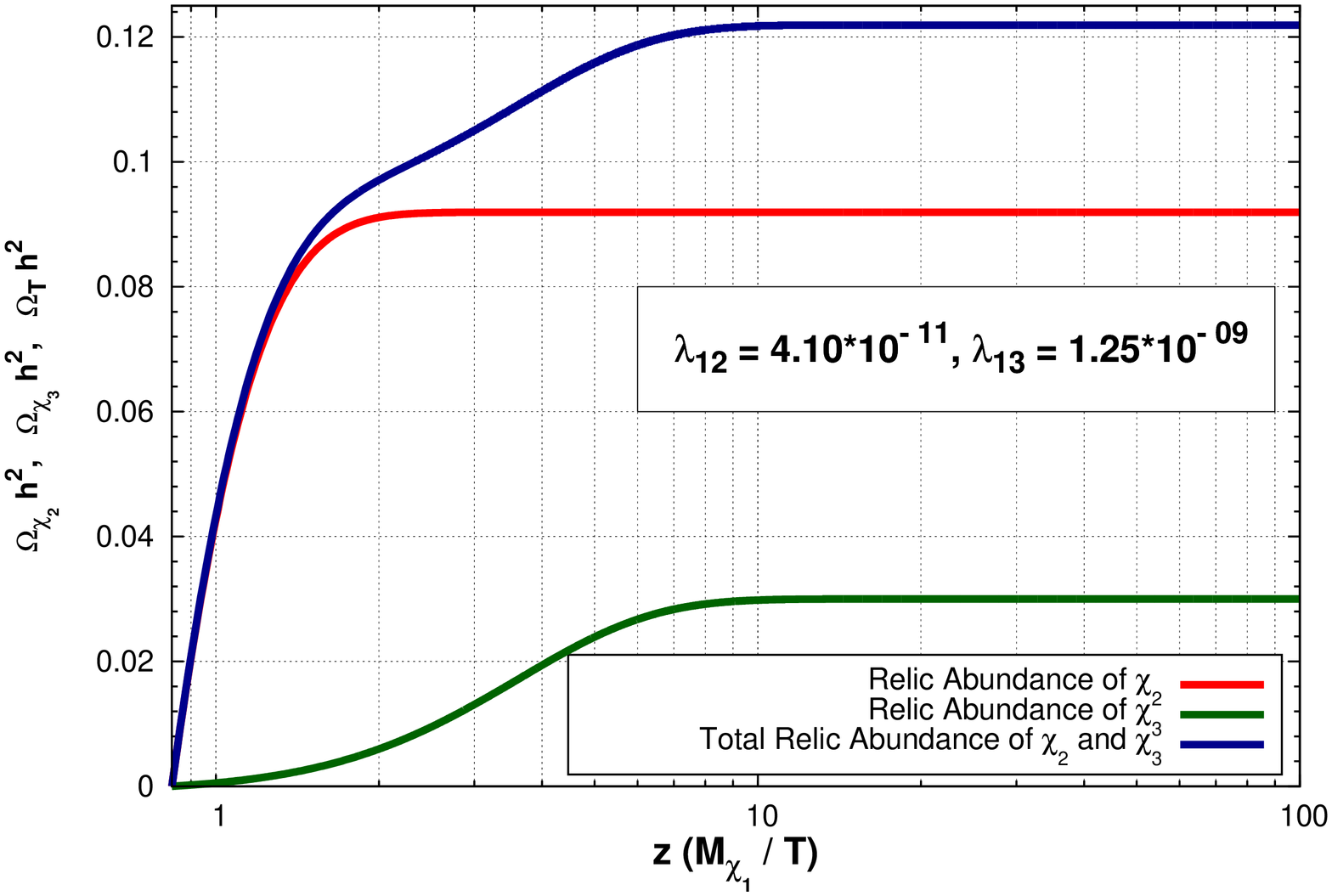}}
\subfigure[$\Omega_{\chi_{{}_{{}_2}}} < \Omega_{\chi_{{}_{{}_3}}}$]
{\includegraphics[height=4.15cm,width=4.5cm,angle=-90]{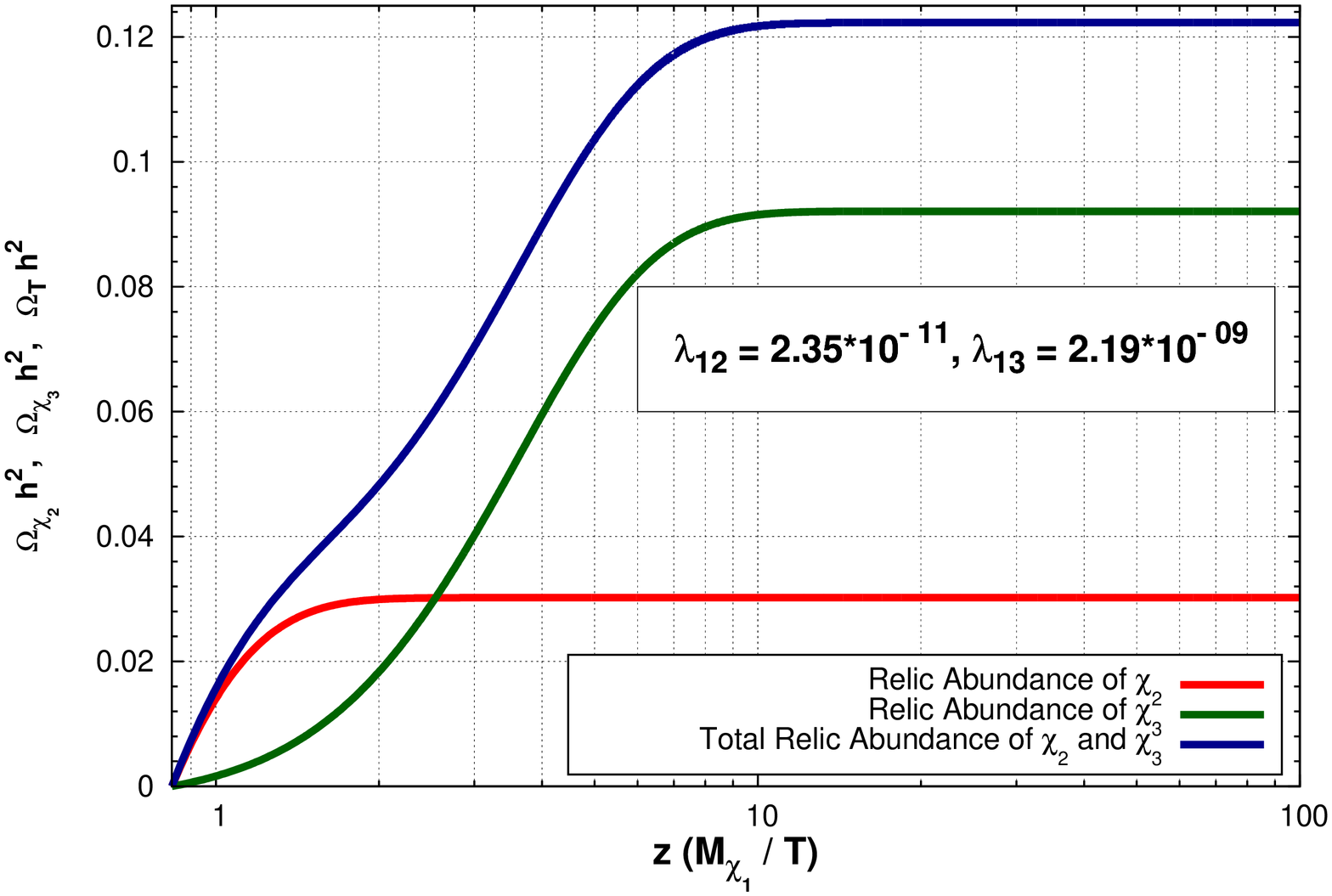}}
\caption{Variation of relic densities of both the dark matter candidates
with $z$.}
\label{relic-density-plot}
\end{figure}

\begin{figure}[h!]
\centering
\includegraphics[height=8cm,width=5cm,angle=-90]{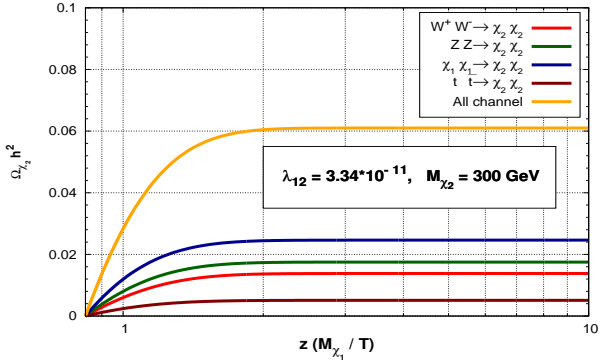}
\caption{Contributions of different production channels to the relic density
of a 300 GeV $\2$ .}
\label{relic-density-chi2}
\end{figure}

We take as a boundary condition the vanishing of $Y_{\jj}$ at
the EW phase transition ($z\sim 0.83$). Figures \ref{relic-density-plot}(a)
and \ref{relic-density-plot}(b)
show the variation of the relic densities of both DM candidates $\4$ with $z$
for different values of $\lambda_{12}$, $\lambda_{13}$ which are $\sim 10^{-9}-10^{-11}$.
Such strengths are needed to keep $Y_{\4}$ small enough to generate the right
DM relic density ($\Omega_{\rm DM}h^2$) at the present epoch. Appropriate values have been
chosen for $\lambda_{12}$, $\lambda_{13}$ depending on whether $\2$ or $\3$
is the dominant DM component. Starting with null values, $Y_{\4}$ are
seen to rise as more and more DM is produced from
the decay/self-annihilation of SM particles. They eventually saturate
to respective particular values at $z\sim 10$ corresponding to a
temperature $T\sim 12$ GeV of the Universe, depending on the
particular values of $\lambda_{12}$, $\lambda_{13}$. These saturation
values together need to satisfy the PLANCK \cite{Ade:2013zuv}
68\% c.l. constraint $0.1172\leq \Omega_{\rm DM} h^2 \leq 0.1226$.
Contributions from individual pair production
channels of $\2$ towards $\Omega_{\2}h^2$ are graphically
shown in Fig.\ref{relic-density-chi2} with chosen parameters
given in its legend: blue line for $\1$, green line for $Z$,
red line for $W$ and brown line for t-quarks, the last being somewhat
less in magnitude. The total relic density of $\2$ (yellow line)
saturates around 0.06 which is half the total DM relic density
($\Omega_{\rm T} h^2$) of today, cf. Ref. \cite{Ade:2013zuv}.
The remainder comes from $\3$.

The allowed ranges of $\lambda_{12}$, $\lambda_{13}$, $\lambda_{23}$,
$\theta_{12}$, $\theta_{13}$, $\theta_{23}$ are given in
Table \ref{table2}.
\begin{table}[h!]
\begin{center}
\begin{tabular} {|c|c|c|}
\hline
{$\lambda_{12}$} & {$\lambda_{13}$} & {$\lambda_{23}$}\\
 & & \\
\hline
$\sim (1.4 - 4.5)$ &$\sim (0.7 - 2.4)$ & $\sim(0.2-4.7)$\\
$\times10^{-11}$&$\times10^{-9}$ & $\times10^{-6}$\\
\hline
{$\theta_{12}$}&$\theta_{13}$&$\theta_{23}$\\
(rad)&(rad)&(rad)\\
\hline
$\sim(0.001-1.75)$&$\sim (0.063-6.3)$&$\sim (0.08-1.31)$\\
$\times 10^{-25}$&$\times10^{-12}$&$\times10^{-13}$\\
\hline
\end{tabular}
\end{center}
\caption{Allowed ranges of
concerned couplings and mixing angles.
Also, $\alpha \sim (10^{-9}-10^{-8})$ GeV$^2$.}
\label{table2}
\end{table}
Given the chosen values of $M_{\3}$ and $u$, the range
of $\lambda_{23}$ is fixed by the need to avoid a late
time decay of $\2$ via $\2\rightarrow \3 \3$.
The tiny magnitudes of $\theta_{13}$, $\theta_{23}$
and $\theta_{12}$ are required by the constraint of keeping the
off diagonal elements of ${\mathcal{M}}^2$ in (\ref{mass-matrix})
to be very small. Further, the couplings of $\4$ with $\1$,
which are functions of the three $\lambda$'s and the three $\theta$'s
\cite{Biswas:2015sva}, remain sufficiently feeble to keep
the former beyond the reach of DM direct detection
experiments [\cite{Aprile:2012nq}-\cite{Akerib:2013tjd}].
Another point to note is that $\2$ behaves like a feebly interacting
massive particle (FIMP) starting with a vanishing number
density. Its fractional relic density saturates after increasing initially
(cf. Fig.\ref{relic-density-plot}a) as the temperature falls
in the cooling Universe. This is the hallmark of a ``freeze-in"
behaviour \cite{Hall:2009bx}, as contrasted with
that of a WIMP; the relic density of the latter starts
from an equilibrium nonzero value, decreases and then freezes out
at a saturation level. Though much lighter, $\3$ also freezes in
a way similar to that of $\2$ (cf. Fig.\ref{relic-density-plot}b).

We turn next to the $\gamma$-excess observed from
RetII covering the range $2-10$ GeV of the FermiLAT
$\gamma$-energy spectrum. With $M_{\2}\sim 250$ GeV,
$\2$ $-$ on account of its nonzero mixing with the
SM-like Higgs boson $\1$ $-$ decays predominately
into $W^+W^-$. Because of the small $\1\2\2$ coupling,
$\2$ pair-annihilation into the same final state, via s-channel
$\1$ exchange, is a negligible competitor. Ours is the first
model explaining the RetII $\gamma$-excess from the decay $\2\rightarrow W^+W^-$
with $\gamma$-rays coming predominantly out of neutral pions
hadronising from $W^\pm$ decaying into $q \bar{q^{\prime}}$ pairs.
Consider the $\gamma$-flux
from RetII at a line of sight distance $\mathfrak{s}$ and subtending
a solid angle $\Delta \Omega$. The differential distribution is
\begin{eqnarray}
\frac{d\Phi_{\gamma}}{d \Omega d E} &=& \frac{1}{4\pi\,M_{\2}}
~\bar{J}~ \Gamma^\prime_{\chi_{{}_{{}_2}}
\rightarrow {W^+} {W^-}} \frac{d N^{W}_{\gamma}}{dE}\,\,.
\label{gamma-flux}
\end{eqnarray}  
Here $\frac{dN^{W}_{\gamma}}{dE}$ is the energy distribution
of each of the two $\gamma$'s of energy $E$ produced from
the $W$ pair, taken numerically from Ref. \cite{Cirelli:2010xx}.
$\bar{J} = \frac{J}{\Delta \Omega}$ represents an average
of the ``astrophysical factor" $J$ \cite{Bonnivard:2015tta} over the opening
solid angle $\Delta \Omega = 2\pi(1-\cos \alpha_{\rm int})$,
the integration angle $\alpha_{\rm int}$ being $0.5^0$ \cite{Geringer-Sameth:2015lua}.
Further,  
\begin{eqnarray}
J = \int\int \rho(\mathfrak{s}, \Omega) d\mathfrak{s}\,d\Omega\,\,, 
\label{j}
\end{eqnarray}
where $\rho(\mathfrak{s}, \Omega)$ describes the variation
of the local dark matter density in the neighbourhood
of RetII. $J$ has been taken to be $10^{18.8}$ GeV cm$^{-2}$
from Ref. \cite{Bonnivard:2015tta}.
Finally, ${\Gamma^\prime}_{\chi_{{}_{{}_2}}\rightarrow {W}^+ {W}^-}$ is the
product of the partial width for the decay ${\chi_{{}_{{}_2}}\rightarrow {W}^+ {W}^-}$
and the fractional relic density for the component $\2$, i.e. 
${\Gamma^\prime}_{\chi_{{}_{{}_2}}\rightarrow {\rm W} {\rm W}} =  
\frac{\Omega_{\chi_{{}_{{}_2}}}}{\Omega_{\rm T}}
\Gamma_{\chi_{{}_{{}_2}}\rightarrow WW}$. Its occurrence in (\ref{gamma-flux})
is necessitated by the two component nature of our DM. The partial width, mentioned above,
is given in a transparent notation by
\begin{eqnarray}
&&\Gamma_{\2\rightarrow W^+W^-} =
\frac{g^2_{{\!}_{WW\2}}}{64\pi}{M^3_{\2}}
(1-4{M^2_W}{M^{-2}_{\2}})^{1/2}
\times \nonumber \\&&
{M^{-4}_W}(1-4{M^2_W}{M^{-2}_{\2}} +12{M^4_W}{M^{-4}_{\2}})\,
\label{gammaww}
\end{eqnarray}
with the coupling $g_{{\!}_{WW\2}}$ given by
\begin{eqnarray*}
-\frac{2M^2_W}{v}(\sin\theta_{12}\cos \theta_{23} +
\cos \theta_{12}\sin \theta_{23} \sin \theta_{13})\,
\end{eqnarray*}
with
$v=2^{-\frac{1}{4}} G^{-\frac{1}{2}}_{F}$, $G_{F}$ being
the Fermi constant.

\begin{figure}[h!]
\centering
\includegraphics[height=8.0cm,width=8.0cm,angle=-90]{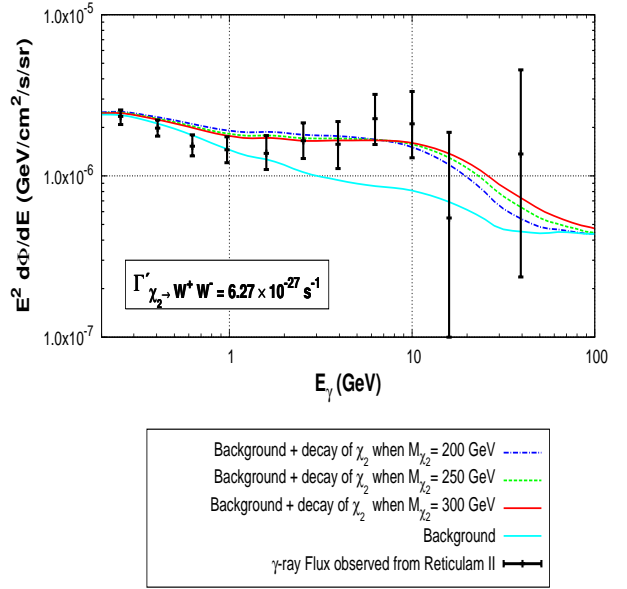}
\caption{Energy distribution of the signal for three different $M_{\2}$'s.}
\label{fermi-result-plot}
\end{figure}

The $\gamma$-flux, computed from (\ref{gamma-flux}), (\ref{j})
and (\ref{gammaww}) for each of the three different values of $M_{\2}$,
is plotted in Fig.\ref{fermi-result-plot} in comparison
with the data points. The background $\gamma$-flux
\cite{Geringer-Sameth:2015lua} (turquoise line) is also shown.
Though the computed plots have been generated
with $\Gamma^{\prime}_{\2\rightarrow W^+W^-}$ fixed at $6.27\times 10^{-27}$ s$^{-1}$,
the fit does not change much, as seen by varying the latter through
$\pm 0.94\times10^{-27}$ s$^{-1}$. In order to produced the above mentioned
range of values of $\Gamma^{\prime}_{\2\rightarrow W^+W^-}$ the soft 
$\mathbb{Z}_2 \times \mathbb{Z}^{\prime}_2$ symmetry breaking parameter
$\alpha$ needs to be in the range 
$10^{-9}\,\,{\rm GeV}^{2}\la \alpha \la 10^{-8}\,\,{\rm GeV}^{2}$.
Clearly, the fit is worse when $M_{\2}$ becomes 200  GeV.
We have not extended our fits to cover $\2$ much beyond 300 GeV
since the production of $\2$ (say from $t\bar{t}$ at the
EW transition temperature $\sim$ 153 GeV) is then cut off
by phase space.
 
\begin{figure}[h!]
\centering
\includegraphics[height=8cm,width=5cm,angle=-90]{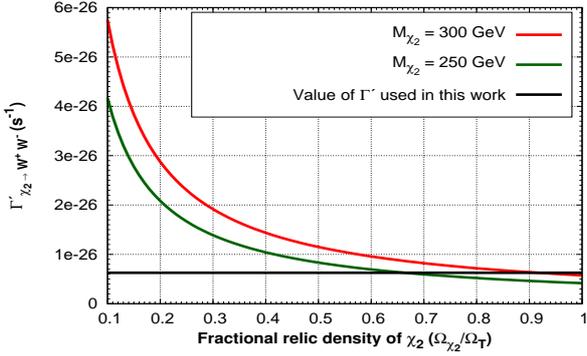}
\caption{${\Gamma^{\prime}}_{\2 \rightarrow W^+ W^-}$ plotted
against $\Omega_{\2}/\Omega_{\rm T}$: AMS-02 upper bound (red for $M_{\2}=300$ GeV
and green for $M_{\2}=250$ GeV) as well as our fixed value (black line).}
\label{ams02}
\end{figure}

Let us discuss indirect constraints on $\Gamma^{\prime}_{\2 \rightarrow W^+ W^-}$
from other observations. First, consider the limit from the positron flux in the
AMS-02 data \cite{ams02data}. Using this data and assuming a single component
DM, Ibarra {\it et al.} \cite{ams02ana} plotted a lower limit (their Fig.3)
on the partial lifetime $\Gamma^{-1}_{{\rm DM} \rightarrow W^+ W^-}$ of the
DM particle decaying into $W^+ W^-$ as a function of the DM mass.
Since we have a two-component DM in our scenario, we need to
consider $\Gamma^{\prime}_{\2 \rightarrow W^+ W^-}$ instead of
$\Gamma_{\2 \rightarrow W^+ W^-}$. (Note that the latter reduces
to the former when $\Omega_{\2}/\Omega_{\rm T}=1$ i.e. one
has a single component DM scenario.) We convert the results of Ref. \cite{ams02ana}
into a plot of the upper limit on $\Gamma^{\prime}_{\2 \rightarrow W^+ W^-}$
as a function of the $\2$ fractional relic density $\Omega_{\2}/\Omega_{\rm T}$
for $M_{\2} = 250$ GeV, 300 GeV. These plots are shown in Fig.\ref{ams02}.
Note that our chosen value of $6.27\times 10^{-27}\,{\rm s}^{-1}$
for $\Gamma^{\prime}_{\2 \rightarrow W^+ W^-}$, made in order to fit the data from RetII,
is below (cf. Fig.\ref{ams02}) the range of this upper bound 
so long as $\Omega_{\2}/\Omega_{\rm T}$ is less than $\sim 0.65\,(0.9)$
for $M_{\2}=250\,{\rm GeV}\,(300\,{\rm GeV})$. Therefore, $\Omega_{\2}$ in
our model is constrained to be less than $\sim 0.65\,(0.9)$ times
the total relic density $\Omega_{\rm T}$ for
$M_{\2}=250\,{\rm GeV}\,(300\,{\rm GeV})$.

\begin{figure}[h!]
\centering
\centering
\subfigure[]
{\includegraphics[height=4.15cm,width=4.5cm,angle=-90]{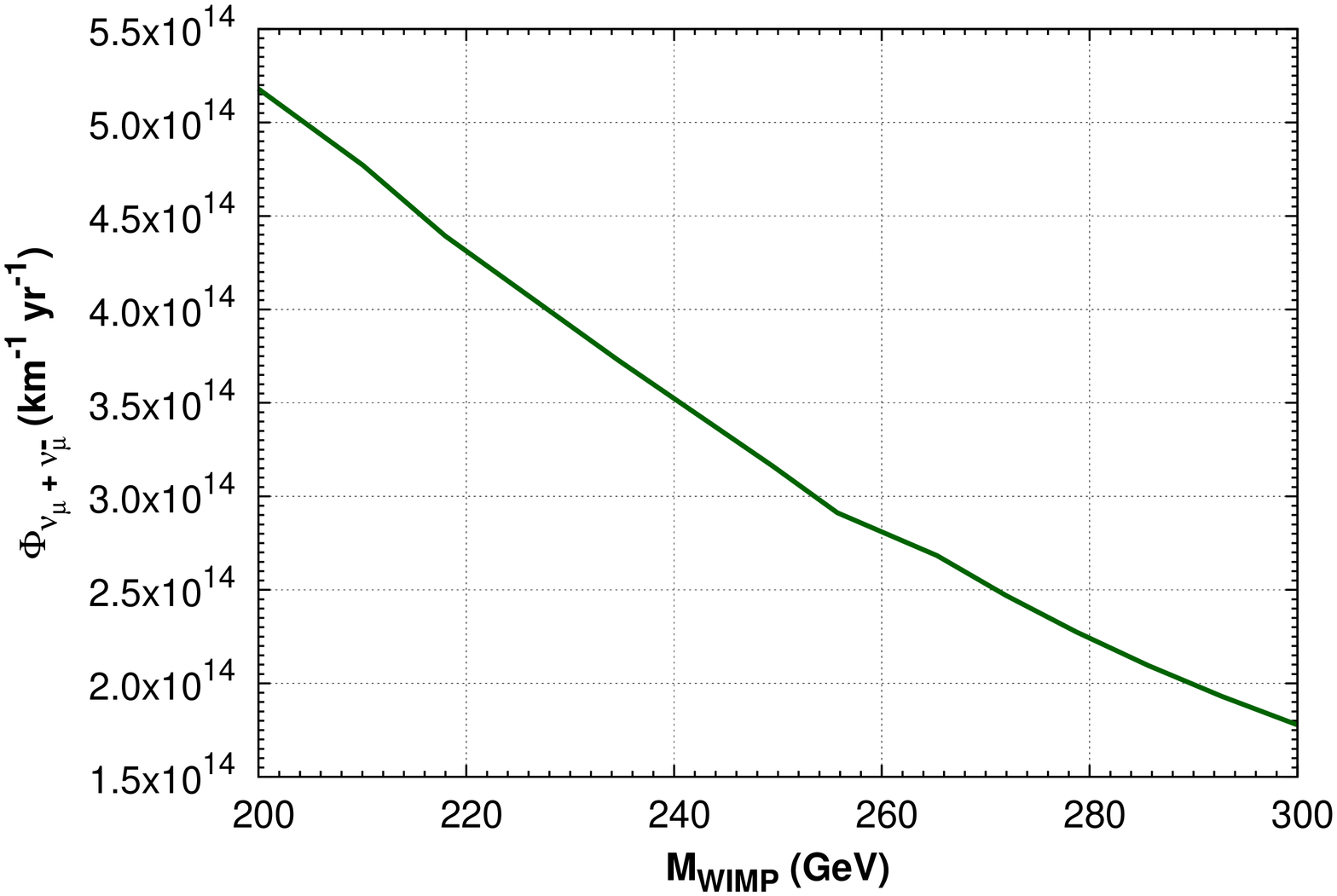}}
\subfigure[]
{\includegraphics[height=4.15cm,width=4.5cm,angle=-90]{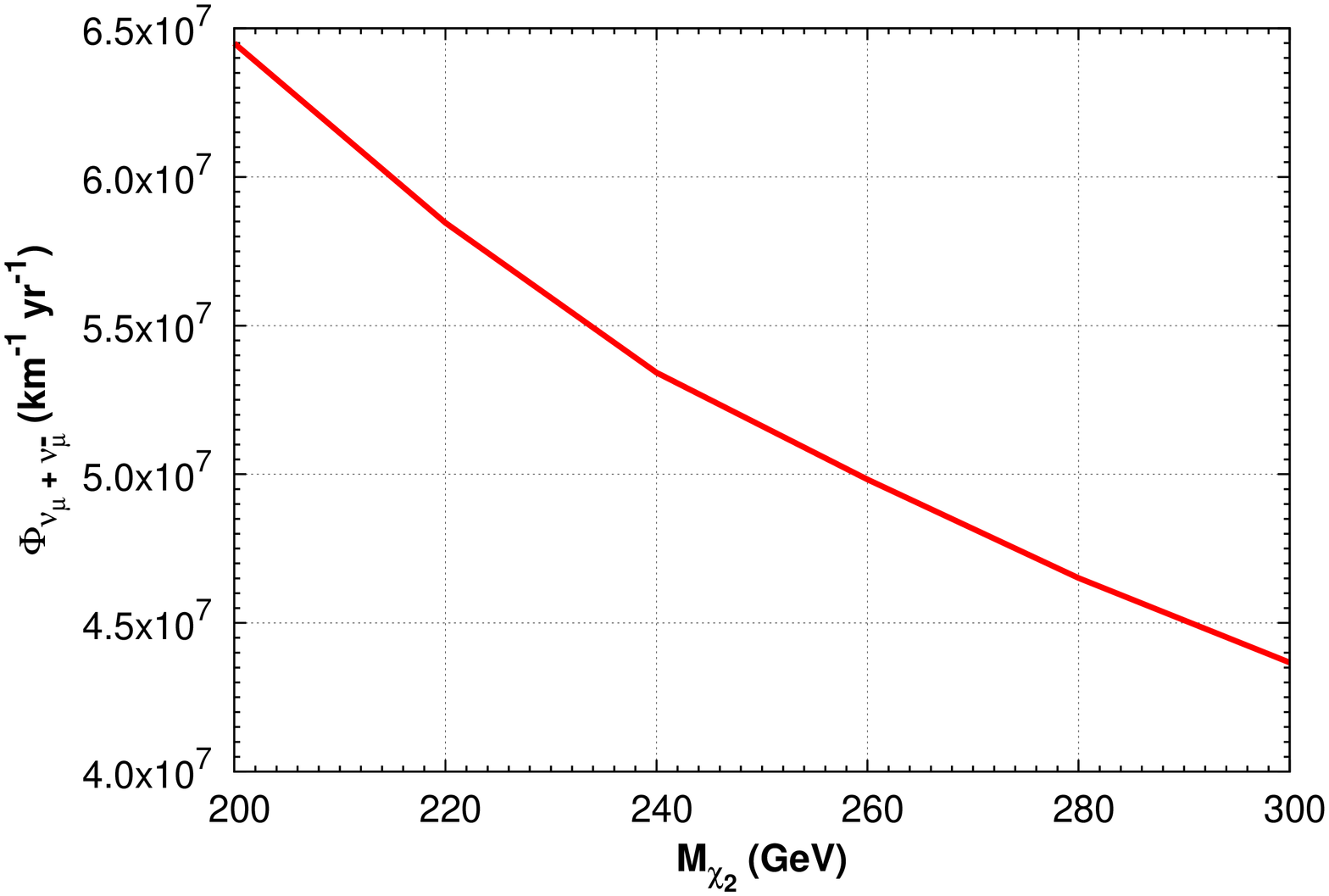}}
\caption{(a) ANTARES upper bound on $\Phi_{\nu_{\mu} + \bar{\nu}_{\mu}}$
(from DM pair-annihilation). (b) $\Phi_{\nu_{\mu} + \bar{\nu}_{\mu}}$
from $\2\rightarrow W^+ W^-$ in our model.}
\label{neutrino}
\end{figure}

\begin{figure}[h!]
\centering
\includegraphics[height=4.15cm,width=4.5cm,angle=-90]{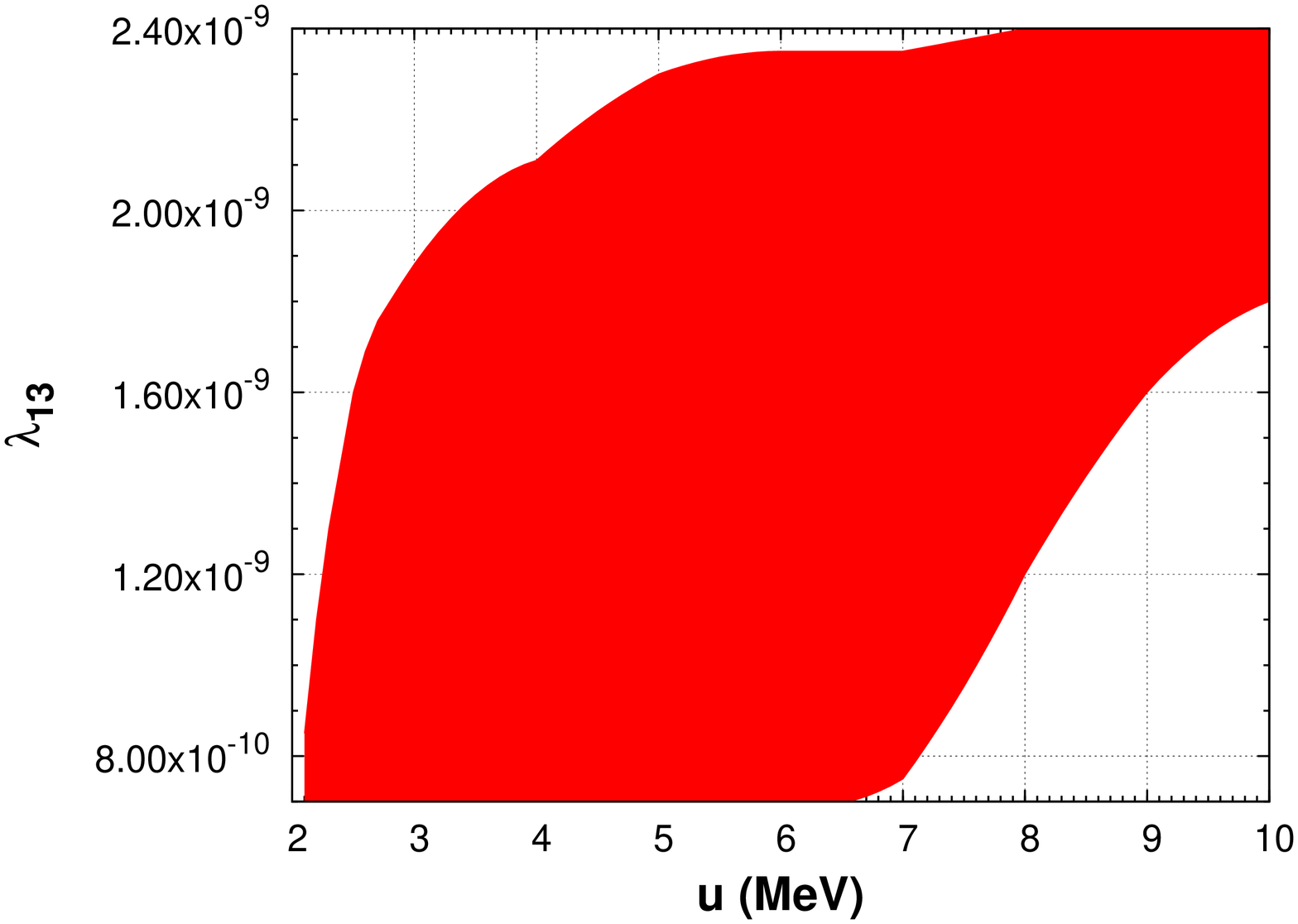}
\includegraphics[height=4.15cm,width=4.5cm,angle=-90]{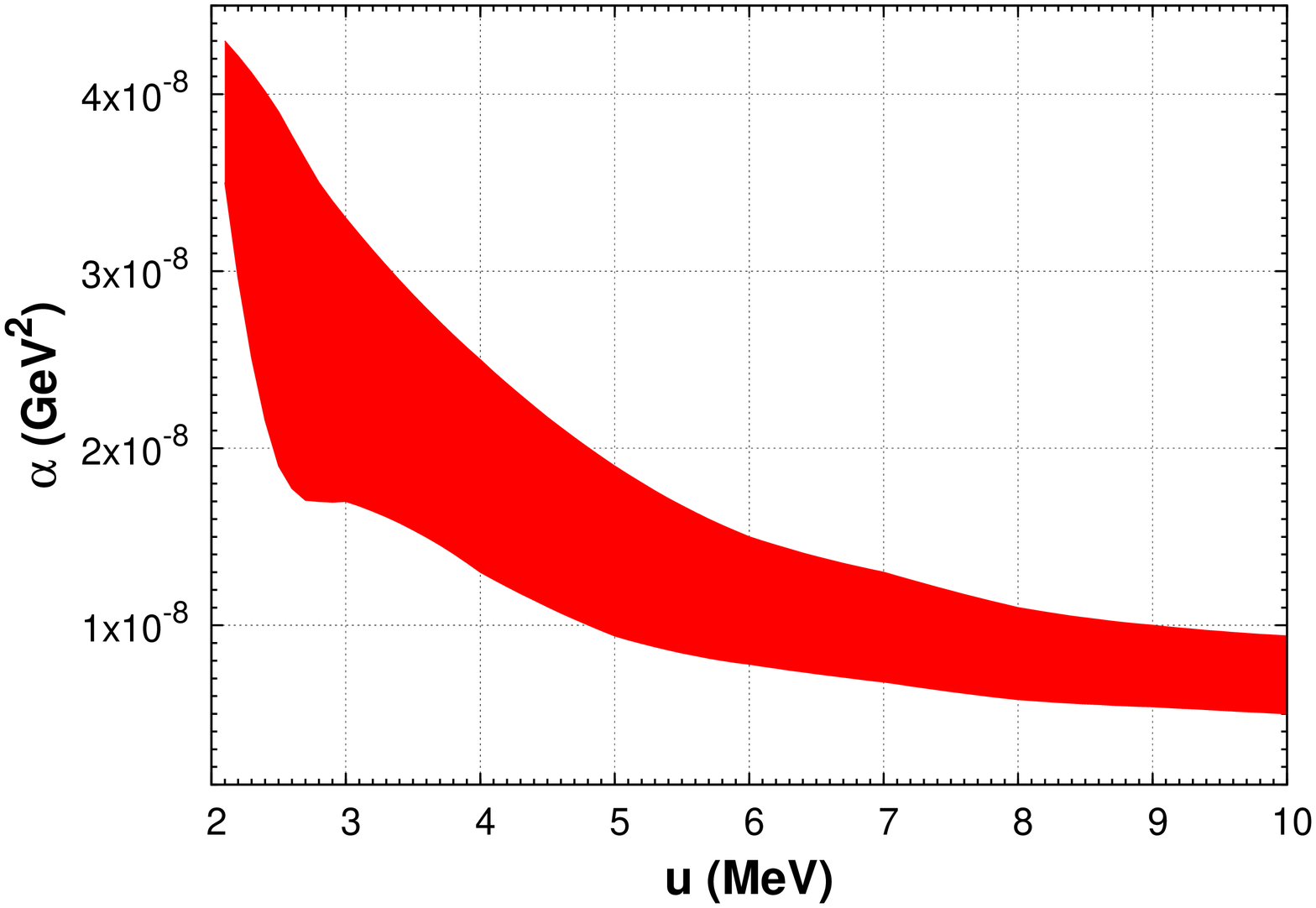}
\caption{Allowed regions in the $u-\lambda_{13}$ (left panel)
and $u-\alpha$ (right panel) planes.}
\label{xray-parameter-plot}
\end{figure}

We next turn to the ANTARES \cite{antares} null result on the
observation of muon neutrinos and antineutrinos from DM processes
at the Galactic Centre. They derived a 90\% c.l. upper bound
on the total flux $\Phi_{\nu_{\mu} + \bar{\nu}_{\mu}}$ as a function of
the mass of the DM particle taking its pair-annihilation
into ${\rm b} \bar{\rm b}$ as the dominant subprocess. This is reproduced
in the left panel of Fig.\ref{neutrino}. In our case the dominant subprocess
is the decay $\2\rightarrow W^+ W^-$. The muon neutrino plus antineutrino
flux from RetII, consequent upon the decays of the $W$'s, is plotted against $M_{\2}$
in the right panel of Fig.\ref{neutrino}. Evidently our flux, being several orders
of magnitude lower, is well within the ANTARES limit. 

The 3.55 keV X-ray line comes from one of the two monoenergetic photons
into which $\3$ decays through its tiny mixing with the SM-like Higgs boson $\1$.
The corresponding modified partial decay width
$\Gamma^{\prime}_{\3\rightarrow \gamma \gamma}
= \Gamma_{\3\rightarrow \gamma \gamma} \frac{\Omega_{\3}}{\Omega_{\rm T}}$
is constrained to be in the range $2.5\times10^{-29}$ s$^{-1}$ $-$
$2.5\times10^{-28}$ s$^{-1}$ in order to fit the observed data. The computation
of $\Gamma_{\3\rightarrow \gamma \gamma}$ is detailed in Ref. \cite{Biswas:2015sva}
and need not to be repeated here. The left (right) panel of Fig.\ref{xray-parameter-plot}
shows the region in the $u-\lambda_{13}$ ($u-\alpha$)
plane allowed by the observational constraints.
The red coloured patch in the left panel is the region compatible with observed
$\gamma$-ray and X-ray fluxes as well as the PLANCK limit on the total DM relic density.
Similar is the case with the patch in the right panel. It is clear from both panels
that those constraints restricts the $\3$ VEV $u$ to $u>2$ MeV. On the other hand,
domain wall constraints [\cite{Biswas:2015sva},\cite{Babu:2014pxa}]
lead to the upper bound $u\leq10$ MeV, mentioned earlier. A noteworthy
fact is that the allowed ranges of the mixing angles $\theta_{12}$, $\theta_{13}$
$-$ given in Table \ref{table2} only from relic density constraints $-$ are
further reduced to
$4.5\times10^{-27} \la \theta_{12}\la 1.67\times10^{-26}$ and
$1.0\times10^{-13} \la \theta_{13}\la 2.75\times10^{-12}$
from the requirement of producing the correct X-ray and $\gamma$-ray
fluxes. The allowed ranges of the other parameters in Table \ref{table2}
remain the same.

In summary, our earlier model \cite{Biswas:2015sva} can fit the analysed
data from RetII, while retaining the explanation for the 3.55 keV X-ray
line $-$ but with substantial modifications. $M_{\2}$ has to be pushed up
to around $250$ GeV. Further, $W^+W^-$ need to replace $\rm{b} \bar{\rm b}$
among the decay products of $\2$ as the primary source of the $\gamma$-excess.
This new seed mechanism requires new Boltzmann equations. They have been formulated
with their consequences quantitatively worked out. 
The compatibility with other indirect constraints has been
checked. The entire picture hangs together.
\acknowledgments
The research of A.B. has been funded by the Department of
Atomic Energy (DAE) of Govt. of India. P.R. has been supported as a
Senior Scientist by the Indian National Science Academy.\\

\end{document}